\documentstyle[12pt,psfig,epsfig]{article}
\newcommand{\beq}{\begin{equation}}
\newcommand{\eeq}{\end{equation}}
\newcommand{\beqarray}{\begin{eqnarray}}
\newcommand{\eeqarray}{\end{eqnarray}}
\def\lsim{\raise0.3ex\hbox{$\;<$\kern-0.75em\raise-1.1ex\hbox{$\sim\;$}}}
\def\gsim{\raise0.3ex\hbox{$\;>$\kern-0.75em\raise-1.1ex\hbox{$\sim\;$}}}
\def\para{\vspace{0.3cm}\noindent}

\begin{document}
\begin{center}
{\large \bf Observability of neutron events above the Greisen-Zatsepin-Kuzmin 
cut-off due to violation of Lorentz invariance} 

\medskip

{Nayantara Gupta \footnote{Email address: nayan@imsc.res.in}}\\  
{\it Institute of Mathematical Sciences,\\
 C.I.T Campus, Taramani, Chennai 600113\\
 INDIA.}  
\end{center}

\begin{abstract}
The clustering of ultra high energy cosmic ray events suggests that they have
originated from compact sources. One of the possible physical mechanisms by
 which ultra high energy nuclei reach the Earth from far away 
astrophysical sources (quasars or BL Lac objects) evading the
 Greisen-Zatsepin-Kuzmin (GZK) cut-off is by violation of Lorentz invariance. 
Assuming that there is violation of Lorentz invariance, we calculate the 
expected number of neutron events from some of the EGRET sources 
(including $\gamma$-ray loud BL Lac objects) which can be correlated in 
direction with ultra high energy cosmic ray events observed by AGASA above 
energy $4\times10^{19}$eV.
 We present in this paper what AGASA should see in future if violation of
Lorentz invariance is responsible for the propagation of ultra high energy
cosmic rays having energies above the GZK cut-off when there is a correlation 
of EGRET sources with the ultra high energy cosmic ray events.    
   
\end{abstract} 

\vskip 2cm
Observational data on ultra high energy cosmic ray events have been recorded
 by different experiments like Fly's Eye \cite{fly1,abu1}, 
Akeno Giant Air Shower Array (AGASA) \cite{aga},
 Yakutsk experiment \cite{yak}, Haverah Park \cite{hav}, 
Volcano Ranch experiment \cite{vol} and Sydney 
University Giant Air-shower Recorder (SUGAR) \cite{sug}.
 The composition of the
ultra high energy cosmic rays (UHECR) has been studied in \cite{fly2,hay1}. 
Data from Fly's Eye experiment \cite{fly2} suggest that the chemical
 composition is dominated by heavy nuclei up to ankle $(10^{18.5})$eV and 
thereafter by lighter component like protons. The AGASA data \cite{hay1} 
suggest a mixed composition of both protons and heavier nuclei. The present 
experiments do not give us enough information about the chemical composition of 
primary UHECR. This 
topic has been addressed in \cite{daw}. If the UHECR are nuclei, then 
what could be their origin and how they propagate in the extragalactic 
magnetic field are at present exciting fields of study. References
\cite{stecker,ahn,sigl1} discuss different suggestions on these issues.    
 
\para 
The Greisen-Zatsepin-Kuzmin(GZK) cut-off \cite{gr} will prevent nucleons of 
energy greater than $4\times10^{19}$eV from travelling more than about 50 Mpc,
 but most of the suitable astrophysical acceleration sites are located at
 greater distances. AGASA experiment \cite{aga} has recorded about 8 events 
above $10^{20}$eV,  whereas data recorded by HiRes fluorescence detector 
\cite{jui} seem to be in agreement with the GZK-cutoff \cite{bahcall}. Since 
currently the experimental data are not convincing enough for accepting the
 presence of the GZK cut-off, theoretical suggestions for observability of 
cosmic rays above the GZK cut-off 
remain of interest to us. The Physics involving the formation and annhilation
 of topological defects (TDs) has been studied as a possible explanation of 
observing UHECR. 
However TD annhilation has unique observational consequences such as, copious
 production of particles like neutrinos and $\gamma$-rays \cite{sigl3}. 
The experimental signatures of the Z-burst scenario has been discussed in 
\cite{weiler}. Gamma ray bursts have also been identified as a possible
 explanation of observing UHECR above the GZK cut-off \cite{eli,vietri}.

\para
If the highest energy cosmic rays have a neutral component then that will come
to the Earth from the direction of its source. Elbert and Sommers \cite{elb}
 identified radio-loud quasar 3C147 as an ideal source within $10^{\circ}$ of
 the highest energy event observed by Fly's Eye experiment \cite{fly1} at 
energy $3.2\times10^{20}$eV. Later another quasar PG0117 + 213 was identified 
by Biermann \cite{bier} within
the error cone of the second highest energy event observed by AGASA \cite{hay2}.
In the last few years a lot of work has been done by different groups 
\cite{farrar2,tiny,gorb,vir} in correlating UHECR events with radio
 quasars and BL Lac objects.
It was pointed out by Dubovsky et al. \cite{dub1} that the statistics of 
clustering of ultra high energy cosmic ray events suggests their correlation
 with compact sources.  
In \cite{torr} the authors have pointed out that there is
no significant correlation between quasars or BL Lac objects and ultra high 
energy cosmic ray events. They have used data from Haverah Park 
 \cite{hav} and Volcano Ranch experiments \cite{vol} for their statistical
analysise. 
However, the authors of \cite{tiny} have shown that there is a significant 
correlation between compact sources and observed ultra high energy cosmic ray 
events using the data from AGASA \cite{aga} and Yakutsk \cite{yak} experiments.
The probability of positional coincidence between BL Lac objects and UHECR
events to occur by chance in a random distribution is of the order of $10^{-4}$.
The still unknown data from HiRes experiment \cite{abu1} and the accumulating
data from AUGER experiment \cite{auger} will enable us to carry out
statistical analysis with a bigger set of data in the near future.    
 
\para
Since, as of now it is not clear whether the UHECR events
are statistically uncorrelated with compact sources at high redshifts, the
physical mechanisms by which neutral component of UHECR may come to the Earth
evading the GZK cut-off from high redshifts are still exciting fields of
study for us. In the Z-burst scenario \cite{farg} or in models of hypothetical
 ``immune messengers'' \cite{chung} neutral particles can come to the Earth 
from the direction of the source evading the GZK cut-off. It has been suggested
by Coleman and Glashow \cite{col} that if there is violation of Lorentz 
invariance (VLI) then protons and neutrons of energies above the GZK cut-off
will reach the Earth without interacting with the cosmic microwave
background.
In the present work we show that one can test whether VLI 
 is responsible for the observability of the UHECR events above the GZK 
cut-off using the data from AGASA experiment and certainly with the data
from AUGER \cite{auger} experiment in future. We have predicted what AGASA
 should see in future if VLI is the underlying physical mechanism for 
propagation of cosmic rays with energies above the GZK cut-off from compact 
sources at high redshifts.

\para
We consider the ultra high energy cosmic ray events observed by AGASA above the
 GZK cut-off. Among this set of events we find that some of them can be correlated
in direction with EGRET sources from the third EGRET catalog \cite{egret}.
 In \cite{gorb} the authors have suggested that the sources of
 UHECR are high energy peaked BL Lac objects. Our set of correlated EGRET
 sources also include $\gamma$-ray loud BL Lac objects.
If the radiation energy density is sufficiently high in a source, photo-pion 
production leads to the generation of a sufficient number of neutrons which 
can escape from the system. If there is VLI the
neutrons having an energy above a certain energy will not decay. The energy 
 above which the neutrons become stable depends on the degree of VLI.
In \cite{dub} the authors have obtained a limit on the energy above which
 neutrons become stable, using observational data from the Yakutsk experiment 
\cite{yak}. These neutrons can travel through the cosmic microwave
background with energy above the GZK cut-off when there is VLI and
in that case we expect to detect them from the direction
of the compact source. 
 For clusters of UHECR events we consider sources within $4^{\circ}$ of the
 events and for a single event we correlate with a source within $2.5^{\circ}$.
In our correlated data set we have eight EGRET sources, among which three 
(3EG J0433+2908, 3EG J1052+5718 and 3EG J1424+3734) are 
 confirmed BL Lac objects from the V{\'e}ron 2001 catalog \cite{ver,gorb}.
 Among the remaining five, 3EG J1744-0310 is a quasar \cite{hew}. The four 
unidentified EGRET sources i.e. 3EG J1824+3441, 3EG J1903+0550, 3EG J0429+0337
 and 3EG J0215+1123 may be BL Lac objects that have not yet been confirmed.
Two of these four unidentified EGRET sources are also present in Table 3 of
\cite{gorb}.
\newpage
\vskip 1cm

TABLE I. The EGRET sources correlated with UHECR events above $4\times 10^{19}$
eV in AGASA data
\begin{center}
\begin{tabular}{c c c c c c c c c} 
\hline\hline
3EG J  & EGRET ID & Type of  & $l_{s}$ & $b_{s}$&$z$ & $l$ & $b$ & $E$\\
       &          & Object& (deg) &(deg) & &(deg) &(deg)& $10^{19}eV$\\
(1)    &  (2)&     (3)  & (4)  &  (5)   &  (6) & (7) & (8) & (9)\\
\hline
0433+2908 & AGN&BL Lac&170.5&-12.6& $>0.3$&170.4&-11.2&5.47\\
          &    & 2EG J0432+2910 &     &     &      &     &     &     \\
          &    &              &     &     &     &171.1&-10.8&4.89\\                                          
         
1052+5718&Possible AGN&BL Lac&149.6&54.42&0.144&147.5&56.2&5.35\\
         &           &RGB J1058+564 &     &     &      &    &    &    \\
         &            &             &     &     &     &145.5&55.1&7.76\\
1424+3734&    &BL Lac&63.95&66.92&0.564&68.5&69.1&4.97\\
         &    &  TEX 1428+370  &     &      &     &   &    &     \\

1744-0310& AGN &Quasar &22.19&13.42&1.054&22.8&15.7&4.27\\
      
0215+1123& &       &  153.75&-46.37& &152.9&-43.9&4.2\\
0429+0337& &       &191.4&-29.08&    &191.3&-26.5&6.19\\
1824+3441& &       &62.49&20.14&     &63.5&19.4&9.79\\
1903+0550& &       &39.52&-0.05& &39.9&-2.1&7.53\\

\hline
\hline
\end{tabular}
\end{center}
     
In the first row of TABLE I we note that the redshift of the EGRET source 
3EG J0433+2908 is more than 0.3 \cite{hal}. The fourth and fifth columns 
of TABLE I  give longitudes and latitudes of the sources in Galactic 
coordinates and the seventh and eighth columns display longitudes and latitudes
of the UHECR events in Galactic coordinates. 

\para

We consider the sources of TABLE I whose redshifts are known and calculate the 
number of neutron events from them expected to be detected by AGASA in 30 years. 
We calculate the expected number of neutron events from the source 
3EG J0433+2908 assuming its redshift to be 0.3. In future we will come to
know the redshifts of the other EGRET sources and then it will also be possible
 to calculate the expected number of neutron events from them.
There are both theoretical \cite{man} and observational \cite{raw} reasons to
believe that when proton acceleration is being limited by energy losses, the 
luminosity of the object in very high energy cosmic rays $L_{CR}$ is 
approximately equal to its luminosity in gamma rays. 
The cosmic ray luminosity $L_{CR}$ in 
the energy range $1\times 10^{19}$eV$ < E < 4 \times 10^{20}$eV can be assumed 
to be emitted equally in each decade of energy $E$. In that case, in each decade
 of energy we expect the power emitted in UHECR in the energy range of 
$1\times 10^{19}$eV to $4\times10^{20}$eV to be approximately $L_{CR}/10$.
Let `$A$' be the  area of AGASA detector which is $10^{12} cm^{2}$ \cite{aga}.
Here we mention that the exposure of the AGASA detector is energy independent
in the energy range in which we are interested. One can see the plot of 
exposure against energy of the UHECRs for AGASA detector in \cite{aga1}.
The expected number of neutron events in a time interval $dt_{o}$ in AGASA 
 within the source energy interval of $E_1$ and $E_2$ can be expressed as

\beq
\frac{dN^{n}_{o}}{dt_{o}}= \frac{A}{4\pi d^2}\int_{E_1}^{E_2} \frac { dN^{n}_{s}}
{dE_{s} dt_{s}} dE_{s}
\eeq
where $d$ is the luminosity distance of the source from the Earth. 
$N^{n}_{s}$ and $N^{n}_{o}$ are respectively the number of neutrons emitted at 
the source and observed by the detector. $E_{s}$, $E_{o}$ are the source
 energy and observed energy of the neutrons respectively. If the redshift
of the source is $z$ then $E_{o}=E_{s}/(1+z)$. Similary the correction 
 to be applied to observed time $t_{o}$ due to redshift of the source is
 $t_{o}=t_{s}(1+z)$, where $t_{s}$ is the time at which the neutron is emitted
from the source.

We have assumed there is VLI and that the lower limit of the 
above integration is such that neutrons are stable above this energy.

We define $F^{ob}_{\gamma}$ as the energy received in gamma rays per second per
 $cm^{2}$ on the surface of the Earth from an EGRET source and this quantity
can be calculated  from \cite{egret}.
 $\epsilon_n$ is the efficiency of neutron production in the source.
When the photon density in a source is sufficiently high then the efficiency
of neutron and proton production become comparable near the end of the UHECR
spectrum \cite{anch2}. Hence it is not unreasonable to assume $\epsilon_n=1/2$
in our calculation.
  If we assume the luminosity of the source in gamma rays of energy more than
$100$MeV $L_{\gamma}$ to be a fraction $x$ times the luminosity of the source
 in UHECR $L_{CR}$, then eqn.(1) can be written as

\beq
\frac{dN^{n}_{o}}{dt_{o}}=A \epsilon_n x\frac{F^{ob}_{\gamma}}{10} \int_{E_1}^{E_2}
 \frac{dE_s}{E^2_s}
\eeq
An almost similar procedure has been followed in ref.\cite{anch2} in calculating
number of neutron events from Centaurus A but without the assumption of VLI.
Using eqn.(2) we calculate the expected number of neutron events between
 observed energy intervals of $1\times 10^{20}$eV and $2\times 10^{20}$eV
 in AGASA from the sources of TABLE I whose redshifts are known. 
The second column of TABLE II displays the photon flux in $10^{-8}$ photon
 $cm^{-2} sec^{-1}$ above $100$ MeV energy for the EGRET sources in the first
 column \cite{egret}. The third column shows the photon spectral indices from 
\cite{egret} of the EGRET sources displayed in the first column. The fourth
 and fifth columns present the expected number of neutron events in a time 
interval of $30$ years in AGASA for case (1) $\epsilon_{n}=1/2$, $x=1$
and for case(2) $\epsilon_{n}=1/2$, $x=1/2$ respectively.

\newpage
TABLE II. The number of neutron events between energy $1\times 10^{20}$eV and
$2\times10^{20}$eV expected in AGASA in  30 years from some of the EGRET 
sources of TABLE I. 
\begin{center}
\begin{tabular}{c c c c c } 
\hline\hline
3EG J  &Photon flux & Photon Spectral  & Number of Neutron&Number of Neutron\\
       &above 100 MeV & Index  &  Events in case (1)& Events in case (2)\\
\hline
0433+2908 &22.0 &1.9 & 8.4& 4.2  \\
1052+5718 &6.5 &2.51 & 2.2& 1.1   \\
1424+3734 &16.3 & 3.25 & 3.5& 1.7 \\
1744-0310 &21.9 &2.42 & 4.29& 2.1 \\
\hline\hline
\end{tabular}
\end{center}

In the last 10 years data from AGASA there is no event above $10^{20}$eV energy 
from the direction of the sources of first column of TABLE II. If VLI is the 
underlying mechanism for UHECR propagation then we can expect to see UHECR 
events above $10^{20}$eV from the direction of the sources presented in
 TABLE II in future. 
If we increase the area of the detector or the time of data collection then of 
course we would detect more neutron events than that presented in TABLE II. The 
number of neutron events will increase linearly with increasing area or time
 of data collection by the detector.
Since the area of AUGER is $30$ times larger than AGASA, we can expect such
neutron events in a much shorter interval of time in AUGER. 
\vskip 2cm
{\bf Conclusion}\\
In this paper we have presented how the UHECR data above the GZK cut-off
can tell us whether VLI is the underlying mechanism for the propagation of
these cosmic rays from quasars or BL Lac objects at high redshifts.

\end{document}